\documentclass[12pt,journal,onecolumn,draftclsnofoot]{IEEEtran}
\usepackage{balance}
\usepackage{lscape}
\providecommand{\keywords}[1]{\textbf{\textit{Index terms---}} #1}
\usepackage{amssymb}% http://ctan.org/pkg/amssymb
\usepackage{pifont}% http://ctan.org/pkg/pifont
\newcommand{\cmark}{\ding{51}}%
\newcommand{\xmark}{\ding{55}}%
\usepackage[justification=centering]{caption}
\ifCLASSINFOpdf
  \usepackage[pdftex]{graphicx}

\else

  \usepackage[dvips]{graphicx}
  
\fi

\usepackage{amsmath}
\usepackage{rotating}

\ifCLASSOPTIONcompsoc
  \usepackage[caption=false,font=normalsize,labelfont=sf,textfont=sf]{subfig}
\else
  \usepackage[caption=false,font=footnotesize]{subfig}
\fi

\usepackage{url}

\usepackage{enumerate}

\usepackage{fancyhdr}
\usepackage{booktabs}
\begin{document}

\title{Performance and Programming Effort Trade-offs of  Android Persistence Frameworks}

\author{
\IEEEauthorblockN{Zheng ``Jason'' Song, Jing Pu, Junjie Cheng, and Eli Tilevich}\\
\IEEEauthorblockA{Software Innovations Lab\\
Virginia Tech\\
Virginia, Blacksburg 24061\\
Email: \{songz, pjing83, cjunjie, tilevich\}@vt.edu}
}

\maketitle

% page numbering
\thispagestyle{fancy} 
\cfoot{\thepage}
\renewcommand{\headrulewidth}{0pt}
\renewcommand{\footrulewidth}{0pt} 
\pagestyle{fancy}
\cfoot{\thepage}

\begin{abstract}
A fundamental building block of a mobile application is the ability to persist program data between different invocations. Referred to as \emph{persistence}, this functionality is commonly implemented by means of persistence frameworks. Without a clear understanding of the energy consumption, execution time, and programming effort of popular Android persistence frameworks, mobile developers lack guidelines for selecting frameworks for their applications. To bridge this knowledge gap, we report on the results of a systematic study of the performance and programming effort trade-offs of eight Android persistence frameworks, and provide practical recommendations for mobile application developers. 
\end{abstract}
\IEEEpeerreviewmaketitle

\section{Introduction}
Any non-trivial application includes a functionality that preserves and retrieves user data, both during the application session and across sessions; this functionality is commonly referred to as \textit{persistence}. In persistent applications, relational or non-relational database engines preserve user data, which is operated by programmers either by writing raw database operations, or via data persistence frameworks. By providing abstractions on top of raw database operations, data persistence frameworks help streamline the development process.  

As mobile devices continue to replace desktops as the primary computing platform, Android is poised to win the mobile platform contest, taking the 82.8\% share of the mobile market in 2015 \cite{Androidshare2015Q2} with more than 1.6 million applications developed thus far \cite{Androidapps2015July}. Energy efficiency remains one of the key considerations when developing mobile applications \cite{jha2011poorly, kwon2013impact, li2014empirical}, as the energy demands of applications continue to exceed the devices' battery capacity. Consequently, in recent years researchers have focused their efforts on providing Android developers with insights that can be used to improve the energy efficiency of mobile applications. The research literature on the subject includes approaches ranging from general program analysis and modeling \cite{hao2013estimating, tiwari1996instruction, dong2010sesame, cohen2012energy} to application-level analysis \cite{sahin2012initial, hindle2012green, kwon2013reducing, pinto2014mining}.

Despite all the progress made in understanding the energy impact of programming patterns and constructs, a notable omission in the research literature on the topic is the energy consumption of persistence frameworks. Although an indispensable building block of mobile applications, these frameworks have never been systematically studied in this context, which can help programmers gain a comprehensive insight on the overall energy efficiency of modern mobile applications. Furthermore, to be able to make informed decisions when selecting a persistence framework for a mobile application, developers have to be mindful of the energy consumption, execution time, and programming effort trade-offs of major persistence frameworks. 

To that end, this paper reports on the results of a comprehensive study we have conducted to measure and analyze the energy consumption, execution time, and programming effort trade-offs of popular Android persistence frameworks. For Android applications, persistence frameworks expose their APIs to the application developers as either object-relational mappings (ORM), object-oriented (OO) interfaces, or key-value interfaces, according to the underlying database engine. In this article, we consider the persistence libraries most widely used in Android applications \cite{mukherjee2014future}.  In particular, we study six widely used ORM persistence frameworks (ActiveAndroid~\cite{activeandroid}, greenDAO~\cite{greendao}, OrmLite~\cite{ormlite}, Sugar ORM~\cite{sugarorm}, Android SQLite~\cite{sqlite}, DBFlow~\cite{DBFlow}), one OO persistence framework (Java Realm~\cite{realm}), and one key-value database operation framework (Paper~\cite{Paper}) as our experimental targets. These frameworks operate on top of the popular SQLite, Realm, or NoSQL database engines. 

In our experiments, we apply these eight persistence frameworks to different benchmarks, and then compare and contrast the resulting energy consumption, execution time, and programming effort (measured as lines of programmer-written code). Our experiments include a set of micro-benchmarks designed to measure the performance of individual database operations as well as the well-known DaCapo H2 database benchmark \cite{blackburn2006dacapo}. To better understand the noticeable performance and programming effort disparities between persistence frameworks, we also introduce a numerical model that juxtaposes the performance and programming efforts of different persistence frameworks. By applying this model to the benchmarks, we generate several guidelines that can help developers choose the right persistence framework for a given application. 

In other words, one key contribution of our study is informing Android developers about how they can choose a persistence framework that achieves the desired energy/execution time/\-programming effort balance. Depending on the amount of persistence functionality in an application, the choice of a persistence framework may dramatically impact the levels of energy consumption, execution time, and programming effort. By precisely measuring and thoroughly analyzing these characteristics of alternative Android persistence frameworks, this study aims at gaining a deeper understanding of the persistence's impact on the mobile software development ecosystem. The specific questions we want to answer are:

\vspace{10pt}
\noindent \textbf{RQ1.}
How do popular Android persistence frameworks differ in terms of their respective features and capabilities?

\vspace{10pt}
\noindent \textbf{RQ2.} as it is realize
What is the relationship between the persistence framework's features and capabilities and the resulting execution time, energy efficiency, and programming effort?

\vspace{10pt}
\noindent \textbf{RQ3.}
How do the characteristics of an application's database functionality affect the performance of persistence frameworks?

\vspace{10pt}
\noindent \textbf{RQ4.}
Which metrics should be measured to meaningfully assess the appropriateness of a persistence framework for a given mobile application scenario?

To answer \textbf{RQ1}, we analyze the documentation and implementation of the persistence frameworks under study to compare and contrast their features and capabilities. To answer \textbf{RQ2} and \textbf{RQ3}, we measure each of the energy, execution time, and programming effort metrics separately, compute their correlations, as well as analyze and interpret the results. To answer \textbf{RQ4}, we introduce a numerical model, apply it to the benchmarks used for the measurements, and generate several recommendation guidelines.

Based on our experimental results, the main contributions of this article are as follows:
\begin{enumerate}[1]
\item To the best of our knowledge, this is the first study that empirically evaluates the energy, execution time, and programming effort trade-offs of popular Android persistence frameworks. 

\item Our experiments consider multifaceted combinations of factors which may impact the energy consumption and execution time of persistence functionality in real-world applications, which include persistence operations involved, the volume of persisted data, and the number of transactions. 

\item Based on our experimental results, we offer a series of guidelines for Android mobile developers to select the most appropriate persistence framework for their mobile applications. For example, ActiveAndroid or OrmLite fit well for applications processing relatively large data volumes in a read-write fashion. These guidelines can also help the framework developers to optimize their products for the mobile market. 

\item We introduce a numerical model that can be applied to evaluate the fitness of a persistence framework for a given application scenario. Our model considers programming effort in addition to execution time and energy efficiency to provide insights relevant to software developers. 
\end{enumerate}

The rest of this paper is organized as follows. Section \ref{sec:relatedworks} summarizes the related prior work. Section \ref{sec:background} provides the background information for this research. Section \ref{sec:experiment} describes the design of our experimental study. Section \ref{sec:results} presents the study results and interprets our findings. Section \ref{sec:rec} presents our numerical model and offers practical guidelines for Android developers. 
Section \ref{sec:threats} discusses the threats to internal and external validity of our experimental results. Section \ref{sec:conclusion} concludes this article. 

This work extends our previous study of Android persistence frameworks, published in IEEE MASCOTS 2016 \cite{jing16}. This article is a revised and extended version of that paper. In particular, we describe the additional research we conducted, which now includes: 1) a comprehensive analysis of the studied frameworks' features, 2) the measurements and analysis for two additional persistence frameworks, 3) a study of the relationship between database-operations, execution time, and energy consumption, based on our measurements, and 4) a novel empirical model for selecting frameworks for a given set of development requirements. 

%To cover this new technical material, we had to substantially enhance the content of one section and add three major sections to the manuscript. Specifically, as far as the new technical material is concerned, in Section \ref{sec:relatedworks}, we include a summary of the related state of the art in studying the energy consumption of mobile computing. In Section \ref{sec:background}, we compare and contrast the features and capabilities of the eight studied persistence frameworks. In Section \ref{sec:results}, we now evaluate the results of our study more comprehensively than in our prior work. Finally, in Section \ref{sec:rec}, we present our numerical model for use by Android application developers. 

\section{Related Work}\label{sec:relatedworks}
This section discusses some prior approaches that have focused on understanding the execution time, energy efficiency, and programming effort factors as well as their interaction in mobile computing. Multiple prior studies have focused on understanding the energy consumption of different mobile apps and system calls, including approaches ranging from system level modeling of general systems~\cite{tiwari1996instruction} and mobile systems~\cite{dong2010sesame}, to application level analysis~\cite{banerjee2014detecting} and optimization~\cite{kwon2013reducing}. For example, Chowdhury et al. study the energy consumed by logging in Android apps~\cite{chowdhury2017exploratory}, as well as the energy consumed by HTTP/2 in mobile apps~\cite{chowdhury2016client}. Liu et al. \cite{liu2016understanding} study the energy consumed by wake locks in popular Android apps. Many of these works make use of the Green Miner testbed \cite{hindle2014greenminer} to accurately measure the amount of consumed energy.
As an alternative, the Monsoon power meter\cite{monsoon} is also used to measure the energy consumed by various Android APIs~\cite{li2014empirical,linares2014mining}. In our work, we have decided to use the Monsoon power meter, due to the tool's ease of deployment and operation.

Many studies also focus on the impact of software engineering practices on energy consumption. For example, Sahin et al. \cite{sahin2012initial} study the relationship between design patterns and software energy consumption. Hindle et al. \cite{hindle2014greenminer} provide a methodology for measuring the impact of software changes on energy consumption. Hasan et al.~\cite{hasan2016energy} provide a detailed profile of the energy consumed by common operations performed on the Java \texttt{List}, \texttt{Map}, and \texttt{Set} data structures to guide programmers in selecting the correct library classes for different application scenarios. Pinto et al. \cite{pinto2014mining} study how programmers treat the issue of energy consumption throughout the software engineering process.

The knowledge inferred from the aforementioned studies of energy consumption can be applied to optimize the energy usage of mobile apps, either by guiding the developer \cite{cohen2012energy, pathak2012energy} or via automated optimization~\cite{li2016automated}. As discovered in \cite{manotas2016empirical}, mobile app developers tend to be better tuned to the issues of energy consumption than developers in other domains, with a large portion of interviewed developers taking the issue of reducing energy consumptions into account during the development process. 

Local databases are widely used for persisting data in Android apps \cite{lyu2017empirical}, and the corresponding database operation APIs are known to be as ``energy-greedy''\cite{linares2014mining}. Several persistence frameworks have been developed with the goal of alleviating the burden of writing SQL queries by hand. However, how these frameworks affect the energy consumption of Android apps has not been yet studied systematically. To bridge this knowledge gap, in this work, we study not only the performance of such frameworks, but also the programming effort they require, with the resulting knowledge base providing practical guidelines for mobile app developers, who need to decide which persistence framework should be used in a given application scenario.

\section{Background}
\label{sec:background}

To set the context for our work, this section describes the persistence functionality, as it is commonly implemented by means of database engines and persistence frameworks. 

The designers of the Android platform have recognized the importance of persistence by including the SQLite database module with the standard Android image as early as the release 1.5. Ever since this module has been used widely in Android applications. According to our analysis of the most popular 550 applications hosted on GooglePlay (25 most popular apps for each category, and 22 categories in all), over 400 of them (73\%) involve interactions with the SQLite module. 

The ORM (object-relational mapping) frameworks have been introduced and refined to facilitate the creation of database-oriented applications \cite{xia2009efficient, cvetkovic2010comparative}. The prior studies of the ORM frameworks have focused mainly on their \emph{execution efficiency} and \emph{energy efficiency}. Meanwhile, Vetro et al. ~\cite{vetro2013definition} show how various software development factors (e.g., design patterns involved, software architecture, information hiding, implementation of persistence layers, code obfuscation, refactoring, and data structure usage) can significantly influence the performance of a software system. In this article, we compare the performance of different persistence frameworks in a mobile execution environment with the goal of understanding the results from the perspective of mobile app developers. 

\vspace{10pt}
\textbf{Android persistence frameworks}
A persistence framework serves as a middleware layer that bridges the application logic with the database engine's operations. The database engine maintains a schema in memory or on disk, and the framework provides a programming interface for the application to interact with the database engine.

As SQLite (the native database of Android) is a relational database engine, most Android persistence frameworks developed for SQLite are ORM/OO frameworks. One major function of such object-relational mapping (ORM) and object-oriented frameworks is to solve the \emph{the object-relational impedance mismatch} \cite{subramanian1999performance} between the object-oriented programming model and the relational database operation model.  

\newpage
\begin{landscape}
\begin{table*}[h]
\scriptsize
\begin{center}
\begin{tabular}{l|l|l|l|l|l|l|l|l}
\toprule
\textbf{Features} & \textbf{Android SQLite} & \textbf{ActiveAndroid} & \textbf{greenDAO} & \textbf{OrmLite} & \textbf{Sugar ORM} & \textbf{Java Realm} & \textbf{DBFlow} & \textbf{Paper}\\ \hline
\textbf{Database Engine} & SQLite & SQLite & SQLite & SQLite & SQLite & Realm & SQLite & NoSQL \\ \hline
\textbf{\begin{tabular}[c]{@{}l@{}}Object Code \\ Generation\end{tabular}} & \xmark & \xmark & \cmark & \xmark & \xmark & \xmark & \xmark  & N.A. \\ \hline
\textbf{Schema Generation} & Manual & Auto & Auto & Manual & Auto & Auto  & Auto  & N.A \\ \hline
\textbf{Data Access Method} & \xmark & \xmark & \cmark & \cmark & \xmark & \xmark & \cmark & N.A. \\ \hline
\textbf{Relationship Support} & \xmark & \begin{tabular}[c]{@{}l@{}}One-To-One,\\ One-To-Many\end{tabular} & \begin{tabular}[c]{@{}l@{}}One-To-One, \\ One-To-Many, \\ Many-to-Many\end{tabular}  & \xmark & One-To-Many & \begin{tabular}[c]{@{}l@{}}One-To-One, \\ One-To-Many, \\ Many-to-Many\end{tabular}  & One-To-Many  & N.A. \\ \hline
\textbf{\begin{tabular}[c]{@{}l@{}}Raw Query \\ Interface Support\end{tabular}} & \cmark & \xmark & \cmark & \cmark & \xmark & \xmark & \cmark & N.A. \\ \hline
\textbf{Batch Operations} & \xmark & \xmark & \begin{tabular}[c]{@{}l@{}}batch insert, \\ batch update, \\ batch delete\end{tabular} & \begin{tabular}[c]{@{}l@{}}batch insert, \\ batch update, \\ batch delete\end{tabular} & \begin{tabular}[c]{@{}l@{}}batch insert, \\ batch update, \\ batch delete\end{tabular} & batch insert  & \begin{tabular}[c]{@{}l@{}}batch insert, \\ batch update, \\ batch delete\end{tabular} & N.A.                                                                           \\ \hline
\textbf{\begin{tabular}[c]{@{}l@{}}Complex Update \\ Support\end{tabular}} & Relational & Relational & Object & Relational & Object & Object & Object & N.A. \\ \hline
\textbf{Aggregation Support} & All & COUNT & COUNT & All & \begin{tabular}[c]{@{}l@{}}COUNT, FIRST, \\ LAST\end{tabular} & \begin{tabular}[c]{@{}l@{}}MAX, MIN, \\ SUM, AVERAGE, \\ FIRST, LAST\end{tabular} & All & N.A. \\ \hline
\textbf{Key/Index Structure} & \begin{tabular}[c]{@{}l@{}}primary key, \\ index, foreign key\end{tabular} & \begin{tabular}[c]{@{}l@{}}integer single primary \\ key, unique, index, \\ foreign key\end{tabular} & \begin{tabular}[c]{@{}l@{}}integer single primary \\ key, unique, index\end{tabular} & \begin{tabular}[c]{@{}l@{}}single primary \\ key, index, \\ foreign key\end{tabular} & \begin{tabular}[c]{@{}l@{}}integer single \\ primary key, \\ unique\end{tabular} & \begin{tabular}[c]{@{}l@{}}string or integer \\ single primary \\ key, index\end{tabular}  & \begin{tabular}[c]{@{}l@{}}primary key, \\ index, foreign key, unique\end{tabular} & N.A.\\ \hline
\textbf{SQL JOIN Support} & \cmark & \cmark & \cmark & \cmark & \xmark & \xmark & \cmark & N.A. \\ \hline
\textbf{Transaction Support} & \cmark & \cmark & \xmark & \cmark & \xmark & \cmark & \cmark & N.A. \\ \hline
\textbf{Caching Support} & \xmark & \cmark & \cmark & \cmark & \xmark & \xmark  & \cmark & \cmark \\ \hline 

\end{tabular}
\end{center}
\caption{Persistence framework feature comparison}
\vspace{-8mm}
\label{table:featurecompare}
\end{table*}
\end{landscape}

We evaluate 8 frameworks: Android SQLite, ActiveAndroid, greenDAO, OrmLite, Sugar ORM, DBFlow, Java Realm, and Paper, backed up by the SQLite, Realm, and NoSQL database engines, which are customized for mobile devices, with limited resources, including battery power, memory, and processor. 

\textbf{Persistence Framework Feature Comparison}
Persistence frameworks differ in a variety of ways, including database engines, programming support (e.g., object and schema auto generation), programming abstractions (e.g., data access object (DAO) support, relationships, raw query interfaces, batch operations, complex updates, and aggregation operations), relational features support (e.g., key/index Structure, SQL join operations, etc.), and execution modes (e.g., transactions and caching). We focus on these features, as they may impact energy consumption, execution time, and programming effort. Table \ref{table:featurecompare} compares these similarities and differences of the persistence frameworks used in our study.

\begin{enumerate}[1]
\item \textbf{Database Engine} Six of the studied persistence frameworks use SQLite \cite{newman2004sqlite}, an ACID (Atomic, Consistent, Isolated, and Durable) and SQL standard-compliant relational database engine. Java Realm framework uses Realm \cite{realm}, an object-oriented database engine, whose design goal is to provide functionality equivalent to relational engines. Paper uses NoSQL\cite{NoSQL}, a non-relational, schema-free, key-value data storage database. Therefore, as we here mainly compare features of relational database, Paper as a non-relational database engine, lacks many of these features. 

\item \textbf{Object Code Generation} Some frameworks feature code generators, which relieve the developer from having to write by hand the classes that represent the relational schema in place.

\item \textbf{Schema Generation} At the initialization stage, persistence frameworks employ different strategies to generate the database schema. Android SQLite requires raw SQL statements to create database tables, while OrmLite provides a special API call. greenDAO generates a special DAO class that includes the schema. The remaining frameworks (excluding Paper) automatically extract the table schema from the programmer defined entity classes.

\item \textbf{Data Access Method} DAO (Data Access Object) is a well-known abstraction strategy for database access that provides a unified object-oriented, entity-based persistence operation set (insert, update, delete, query, etc.). greenDAO, Sugar ORM, and OrmLite provide the DAO layer, while Android SQLite adopts a relational rather than DAO database manipulating interface. ActiveAndroid, DBFlow, and Java Realm provide a hybrid strategy---both DAO and SQL builder APIs.

\item \textbf{Relationship Support} The three relationships between entities are one-to-one, one-to-many, and many-to-many. greenDAO and Java Realm support all three relationships. ActiveAndroid lacks support for Many-to-Many, while Sugar ORM and DBFlow only support One-To-Many. Android SQLite and OrmLite lack support for relationships, requiring the programmer to write explicit SQL join operations.

\item \textbf{Raw Query Interface Support} Raw queries use naive SQL statements, thus deviating from pure object-orientation to execute complex database operations on multiple tables, nesting queries and aggregation functions. Android SQLite, greenDAO, OrmLite, and DBFlow---all provide this functionality.  

\item \textbf{Batch Operations} Batch operations commit several same-type database changes at once, thus improving performance. greenDAO, OrmLite, Sugar ORM, and DBFLow provide batch mechanisms for insert, update and delete. Java Realm provides batch inserts only, and the remaining two frameworks lack this functionality.

\item \textbf{Complex Update Support} Typically there are two kinds of database update operations: update columns to given values, or update columns based on arithmetic expressions. Android SQLite and ActiveAndroid can only use raw SQL manipulation interface to support expression updates. greenDAO, Sugar ORM, Java Realm, and DBFlow support complex updates via entity field modification. OrmLite is the only framework that provides both the value update and expression update abstractions.

\item \textbf{Aggregation Support} Aggregating data in a relational database enables the statistical analysis over a set of records. Different frameworks selectively implement aggregation functionality. Android SQLite, OrmLite, and DBFlow support all of the aggregation functions via a raw SQL interface. Java Realm and Sugar ORM provide an aggregation subset in the entity layer. ActiveAndroid and greenDAO support only the COUNT aggregation.

\item \textbf{Key/Index Structure}
Key/Index structure identifies individual records, indicates table correlations, and increases the execution speed. Android SQLite and DBFlow fully support the database constraints---single or multiple primary keys (PK), index and foreign key (FK). ActiveAndroid supports integer single PK, unique, index, and FK. greenDAO supports integer single PK, unique, index. OrmLite supports single PK, index, and FK. Sugar ORM supports integer single PK and unique. Java Realm supports string or integer single PK, and index.

\item \textbf{SQL JOIN Support}
SQL JOIN clause combines data from two or more relational tables. Android SQLite and DBFlow only support raw JOIN SQL. ActiveAndroid incorporates JOIN in its object query interface. DAOs of greenDAO and OrmLite provide the JOIN operation. Sugar ORM and Java Realm lack this support.

\item \textbf{Transaction Support}
Transactions perform a sequence of operations as a single logical execution unit. All the studied engines with the exception of greenDAO, Sugar ORM, and Paper provide full transactional support.

\item \textbf{Cache Support} OrmLite, ActiveAndroid, greenDAO, DBFlow and Paper support caching. They provide this advanced feature to maintain persisted entities in memory to speed-up future accesses, at the cost of extra processing required to initialize the cache pool. 

\end{enumerate}

%\textbf{Android Concurrency}
%All of the studied persistence frameworks support multiple-threaded execution. To abstract away the low-level complexities of thread programming, we use the Executor pattern in our DaCapo benchmark. An important question is whether single threaded applications can use multiple cores. We determined that for ARM architecture processors Android applications can indeed benefit from multi-core execution without starting multiple threads. The scheduler divides execution tasks into multiple pieces, distributed to different cores.

\section{Experiment Design}
\label{sec:experiment}
In this section, we explain the main decisions we have made to design our experiments. In particular, we discuss the benchmarks, the measurement variables, and the experimental parameters.

%\begin{figure}[!t]
%\centering
%\includegraphics[width=2.5in]{android_device_architecture.png}
%\caption{Experimental Android Device Architecture}
%\label{fig:devicearchitecture}
%\end{figure}

\subsection{Benchmark Selection}
DaCapo H2 \cite{blackburn2006dacapo} is a well-known Java database benchmark that interacts with the H2 Database Engine via JDBC. This benchmark manipulates a considerable volume of data to emulate bank transactions. The benchmark includes 1) a complex schema and non-trivial functionality, obtained from a real-world production environment. The database structure is complex (12 tables, with 120 table columns and 11 relationship between tables), while the database operations simulate the running of heavy-workload database-oriented applications; 2) complex database operations that require: batching, aggregations, and transactions. 

Since DaCapo relies heavily on relational data structures and operations, we replace H2 with two relational database engines, SQLite or Realm, to adapt this benchmark for Android. In other words, we evaluate the performance of all persistence frameworks under the DaCapo benchmark except for Paper, which is a non-relational database engine.     

However, using the DaCapo benchmark alone cannot provide performance evaluation of persistence frameworks under the low data volumes with simple schema conditions. To establish a baseline for our evaluation, we thus designed a set of micro benchmarks, referred to as \emph{the Android ORM Benchmark}, which features a simple database schema with few data records. Specifically, this benchmark's database structure includes 2 tables comprising 11 table columns, and a varying small number of data records. Besides, this micro-benchmark comprises the fundamental database operation invocations ``create table'', ``insert'', ``delete'', ``select'', and ``update''. As the database operations in many mobile applications tend to be rather simple, the micro-benchmark's results present valuable insights for application developers.

Note that database operations differ from database operation invocations. The invocations refer to calling the interfaces provided by the persistence framework (e.g., ``insert'', ``select'', ``update'' and ``delete''). However, each invocation can result in multiple database operations (e.g., \texttt{android...SQLiteStatement.executeInsert()}). 

\subsection{Test Suite Implementation}
Our experimental setup comprises a mobile app that uses the selected benchmarked frameworks to execute both DaCapo and the Android ORM Benchmark\footnote{All the code used in our experiments can be downloaded from \url{https://github.com/AmberPoo1/PEPBench}.}. Through this app, experimenters can select benchmarks, parameterize the operation and data volume, as well as select ORM frameworks. The implementation of the test suite was directed by two graduate students, each of whom has more than three years of experience in developing commercial database-driven projects. 

As stated above, the DaCapo database benchmark is designed for relational databases, so it would be non-trivial to reimplement it using Paper, which is based on a non-relational database engine (NoSQL). Therefore, the DaCapo benchmark is only applied to seven persistence frameworks, while the Android ORM Benchmark is applied to all eight persistence frameworks. 

For each benchmark, the transaction logic (e.g., creating bank accounts for DaCapo) is implemented using various ORM frameworks, following the design guidelines of these frameworks. For example, greenDAO, Sugar ORM, and OrmLite provide object-oriented data access methods, so their benchmarks' data operations are implemented by using DAO. On the contrary, Android SQLite adopts a relational rather than DAO database manipulating interface, so its benchmark's data operations are implemented by using SQL builders. ActiveAndroid, DBFlow, and Java Realm provide a hybrid strategy---both DAO and SQL builder APIs. For these frameworks, if the benchmark's operation is impossible or non-trivial to  express using DAO, the SQL builder API is used instead. 

\subsection{Parameters and Variables}
Next, we explain the variables used to evaluate the execution time, energy consumption, and programming effort of the studied persistence frameworks. We also describe how these variables are obtained. 

\begin{itemize}
\item \textbf{Overall Execution Time} The overall execution time is the time elapsed from the point when a database transaction is triggered to the point when it completes.
\item \textbf{Read/Write Database Operation Number} We focus on comparing the Read/Write numbers only for SQLite-based frameworks (ActiveAndroid, greenDAO, OrmLite, Sugar ORM, and DBFlow), as such frameworks use SQLite operation interfaces provided by the Android framework to operate on SQLite database. The write operations include executing SQL statements that are used to ``insert'', ``delete'', and ``update'', while the read operations include only ``select''. When performing the same combination of transactions, the differences in Read/Write number is the output of how different persistence frameworks interpret database operation invocations. The read/write ratio can also impact the energy consumption. The operation numbers are obtained by hooking into the SQLite operation interfaces provided by the Android System Library. For those interfaces provided for a certain type of database operation, we mark them as ``Read'' or ``Write''; for those interfaces provided for general SQL execution, we search for certain keywords (e.g., insert, update, select, and delete), in the SQL strings, and further mark them as ``Read'' or ``Write''. 

\item \textbf{Energy Consumption} The energy consumption can be calculated using the real-time current of the Android device's battery. We use the Monsoon Power Monitor \cite{monsoon} to monitor the current and voltage of the device's battery, as shown in Fig. \ref{fig:mainchannel}. As the output voltage of a smartphone's battery remains stable, only the current and time are required to calculate the energy consumed. 
Equation \ref{eq:energy} is used to calculate the overall energy consumption, where $\bar{I}$ is the average current ($mA$), and $t$ is the time window ($ms$). Micro-ampere-hour is a unit of electric charge, commonly used to measure the capacity of electric batteries.

\begin{equation}
\label{eq:energy}
E = \frac{\bar{I} * t}{3600s/hour}
\end{equation}

\begin{figure}[!ht]
\centering
\includegraphics[width=2.5in]{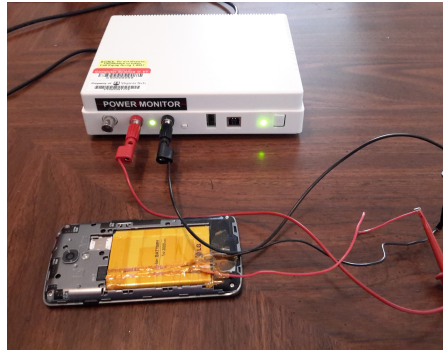}
\caption{Monsoon Mobile Device Power Monitor's main channel measurement connection}
\label{fig:mainchannel}
\end{figure}

The equation shows that the energy consumption is proportional to the execution time, as well as to the current required by the device's hardware (e.g., CPU, memory, network, and hard disk). For the persistence frameworks, the differences of energy consumption reflect not only the execution time differences, but also the different CPU workload and hard disk R/W operations required to process database operations. 

\item \textbf{Uncommented Line of Codes (ULOC)} ULOC reflects the required effort, defined as the amount of code the programmer has to write to use a persistence framework. Currently, ULOC is still the most frequently used way to measure programmer effort in the literature \cite{lavazza2016empirical}. As all the test suites are implemented only in Java and SQL, the ULOC metric is reflective of the relative programming effort incurred by using a framework. 
\end{itemize}

Next, we introduce the input parameters for different benchmarks. For the DaCapo benchmark, we want to explore the performance boundary of different persistence frameworks under a heavy workload. Therefore, we vary the amount of total transactions to a large scale, and record the overall time taken and energy consumed. 

For the micro benchmark, we study the ``initialize'', ``insert'', ``select'', ``update'', and ``delete'' invocations in turn. We change the number of transactions for the last four invocations, so for the ``select'', ``update'' and ``delete'' invocations, the amount of data records also changes. Therefore, the input parameters for the micro benchmark is a set of two parameters:

$\{\textsc{number of transactions}, \textsc{amount of data records}\}$.  

\subsection{Experimental Hardware and Tools}
To measure the energy consumed by a device, its battery must be removed to connect to the power meter's hardware. Unfortunately, a common trend in the design of Android smartphones makes removable batteries a rare exception. The most recently made device we have available for our experiments is an LG LS740 smartphone, with 1GB of RAM, 8GB of ROM and 1.2GHz quad-core Qualcomm Snapdragon 400 processor \cite{singh2014evolution}, running Android 4.4.2 KitKat operating system. Although the device was released in 2014 and runs at a lower CPU frequency than the devices in common use today, the Android hardware design, at least with respect to the performance parameters measured, has not experienced a major transformation, thus not compromising the relevance of our findings.

We execute all experiments as the only load on the device's OS. To minimize the interference from other major sources of energy consumption, we set the screen brightness to the minimal level and turn off the WiFi/GPS/Data modules. As the CPU frequency takes time to normalize once the device exits the sleep mode, we run each benchmark 5 times within the same environment, with the first two runs to warm up the system and wait until the background energy consumption rate stabilizes. The reported data is calculated as the average of the last 3 runs, with the differences of the three runs of the same benchmark being not larger than $5\%$. 

To understand the Dalvik VM method invocations, we use Traceview, an Android profiler that makes it possible to explore the impact of programming abstractions on the overall performance. Unfortunately, only the Android ORM benchmark is suitable for this exploration, due to the Traceview scalability limitations.

\newpage
\begin{figure*}[!ht]
\centering
\includegraphics[width=5.8 in]{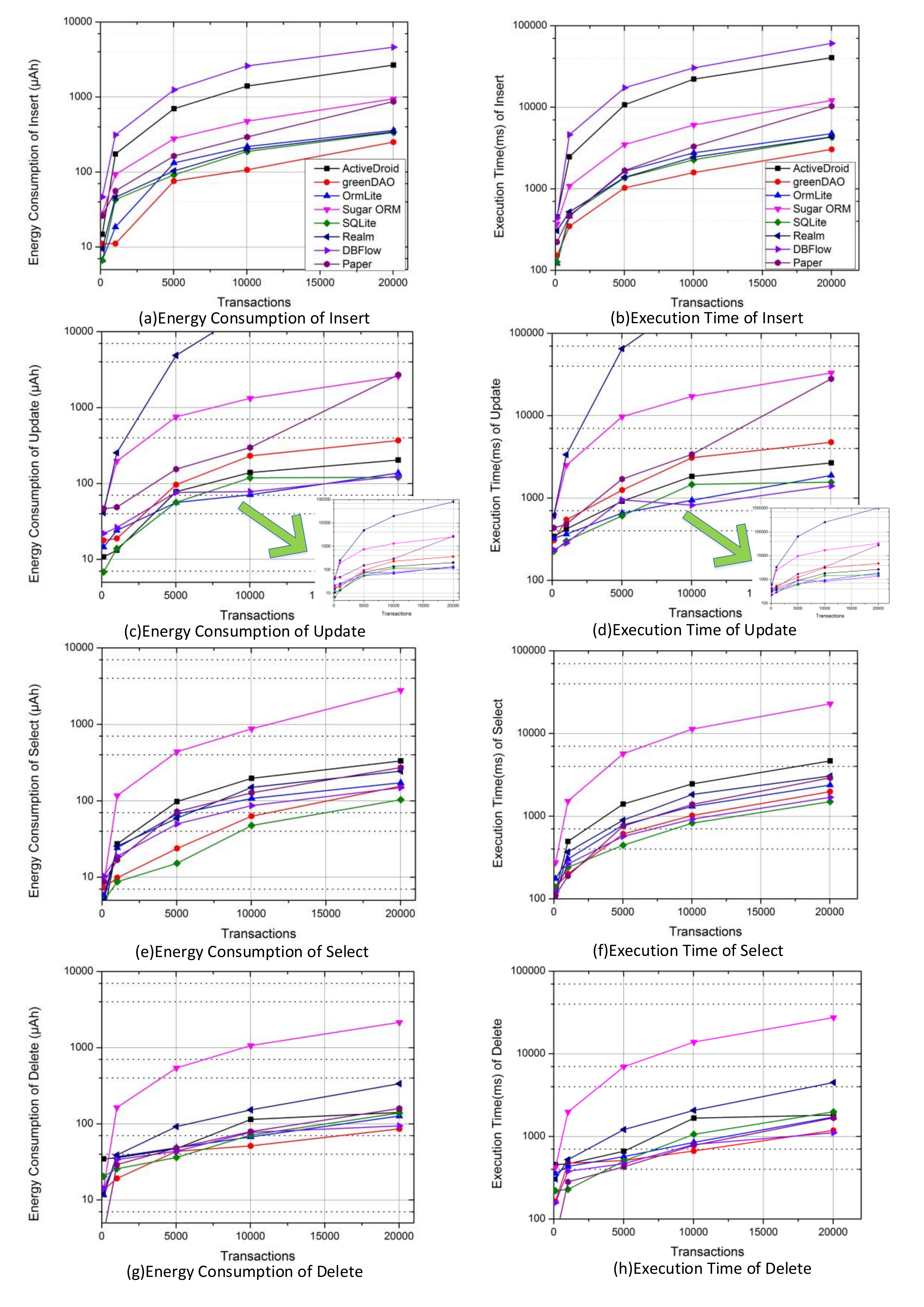}
\vspace{-3mm}
\caption{Energy/Execution time for Android ORM Benchmark \\ with Alternative Persistence Frameworks}
\label{fig:microbenchmark-energytime}
\vspace{-3mm}
\end{figure*}
\newpage

\begin{figure*}[!ht]
\centering
\includegraphics[width=5.5in]{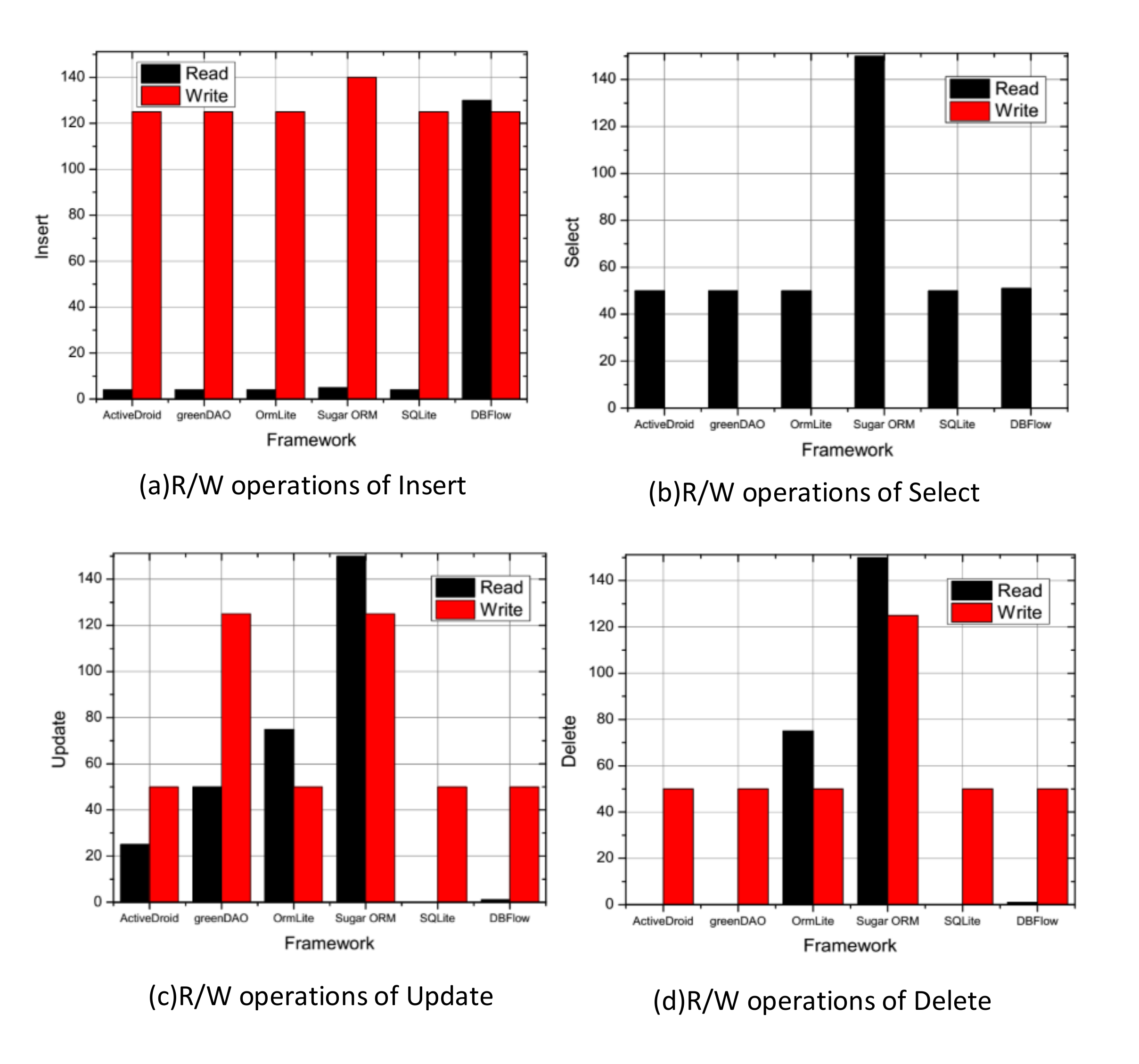}
\vspace{-3mm}
\caption{Read/Write Operations for Android ORM Benchmarks\\ with Alternative Persistence Frameworks}
\label{fig:microbenchmark-dboperations}
\vspace{-3mm}
\end{figure*}

\section{Study Results}
\label{sec:results}
In this section, we report and analyze our experimental results.

\begin{table*}[!ht]
\begin{center}
\begin{tabular}{l|l|l|l|l|l|l|l|l}
\toprule
\textbf{Compared Item}           & \textbf{SQLite} & \textbf{DBFlow} & \textbf{greenDAO} & \textbf{ORMLite} & \textbf{Realm} & \textbf{Paper} & \textbf{Sugar} & \textbf{ActiveAndroid}  \\ \hline
\cline{1-1}
\textbf{ULOC}                    & \textbf{306}    & \textbf{181}    & \textbf{241}      & \textbf{326}     & \textbf{313}   & \textbf{190}   & \textbf{226}   & \textbf{253}            \\ \hline
\cline{1-1}
\textbf{Initialization Ranking}  & \textbf{6}      & \textbf{1}      & \textbf{5}        & \textbf{7}       & \textbf{3}     & \textbf{4}     & \textbf{2}     & \textbf{8}              \\ \hline
\cline{1-1}
\textbf{Insert Ranking}          & \textbf{2}      & \textbf{8}      & \textbf{1}        & \textbf{4}       & \textbf{3}     & \textbf{5}     & \textbf{6}     & \textbf{7}              \\ \hline
\cline{1-1}
\textbf{Update Ranking}          & \textbf{1}      & \textbf{2}      & \textbf{5}        & \textbf{3}       & \textbf{8}     & \textbf{7}     & \textbf{6}     & \textbf{4}              \\ \hline
\cline{1-1}
\textbf{Select Ranking}          & \textbf{1}      & \textbf{2}      & \textbf{3}        & \textbf{4}       & \textbf{5}     & \textbf{6}     & \textbf{8}     & \textbf{7}              \\ \hline
\cline{1-1}
\textbf{Delete Ranking}          & \textbf{4}      & \textbf{2}      & \textbf{1}        & \textbf{3}       & \textbf{7}     & \textbf{6}     & \textbf{8}     & \textbf{5}              \\ \hline
\cline{1-1}
\textbf{Summed up Ranking}       & \textbf{14}     & \textbf{15}     & \textbf{15}       & \textbf{21}      & \textbf{26}    & \textbf{28}    & \textbf{30}    & \textbf{31}             \\ \hline
\cline{1-1}
\end{tabular}
\end{center}
\caption{Comparison of Persistence Frameworks in the Android ORM experiment }
\label{table:microranking}
\vspace{-8mm}
\end{table*}

\subsection{Experiments with the Android ORM benchmark} \label{subsec:5a}

In this group of experiments, we study how the types of operation (insert, update, select, and delete) and the variations on the number of transactions impact energy consumption and execution time with different frameworks using the micro benchmark\footnote{We use the terms \emph{the micro benchmark} and \emph{the Android ORM benchmark} interchangeably in the rest of the presentation.}. The experimental results for each type of persistence operation are presented in \figurename \ref{fig:microbenchmark-energytime} and \figurename \ref{fig:microbenchmark-dboperations}. The first row of \figurename \ref{fig:microbenchmark-energytime}(a)-(b) shows the energy consumption and execution time of the ``insert'' database invocation, and \figurename \ref{fig:microbenchmark-energytime}(c)(d), (e)(f), and (g)(h) show that of the ``update'', ``select'', and ``delete'' database invocations, respectively. \figurename \ref{fig:microbenchmark-dboperations} (a)-(d) show the database read and write operations of these four invocations, respectively. 

The results show that the persistence frameworks differ in terms of their respective energy consumption, execution time, read, and write measurements. Next, we compare the results by operation: 

\textbf{Insert} We observe that DBFlow takes the longest time to perform the insert operation, while ActiveAndroid takes the second longest, with the remaining frameworks showing comparable performance levels. DBFlow performs the highest number of database operations, a measurement that explains its long execution time. Different from other frameworks, DBFlow requires that a database read operation be performed before a database write operation to ensure that the-record-to-insert has not been already inserted into the database. Besides, the runtime trace reveals that interactions with the cache triggered by inserts in ActiveAndroid are expensive, costing 62\% of the overall execution time. By contrast, greenDAO exhibits the shortest execution time, due to its simple but efficient batch insert encapsulation, as shown in Table \ref{table:featurecompare}. 

\textbf{Update} We observe that the cost of the Java Realm update is several orders of magnitude larger than that of other frameworks, especially as the number of transactions grows. Several reasons can explain the high performance costs of the update operation in Java Realm. As one can see in Table \ref{table:featurecompare}, Java Realm lacks support for batch updates. Besides, the update procedure invokes the underlying library method, \texttt{TableView.size()}, which operates on a memory-hosted list of entities and costs more than 98\% of the overall execution time. The execution time of Sugar ORM is also high, due to it having the highest number of read and write operations. Sugar ORM needs to search for the target object before updating it. This search procedure is designed as the recursive \texttt{SugarRecord.find()} method, which costs 96\% of the overall execution time.

\textbf{Select and Delete} For the select and delete operations, we observe that Sugar ORM 1) exhibits the worst performance in terms of execution time and energy consumption; 2) performs the highest number of database operations, as it executes an extra query for each atomic operation. The inefficiency of the select and delete operations in Sugar ORM stems from the presence of these extra underlying operations. However, as discussed above, the bulk of the execution time is spent in the recursive $find$ method. OrmLite, greenDAO, DBFlow, Paper, and Android SQLite show comparable performance levels when executing these two operations.

Table \ref{table:microranking} sums up the rankings of each persistence framework w.r.t. different database operation invocations. We also measure the Uncommented Lines of Code (ULOC) for implementing all the basic database operation invocations for each persistence framework and include this metric in the table. From Table \ref{table:microranking} and our analysis above, we can draw the following conclusions: 
\begin{enumerate}[1]
\item By adding up the rankings of different operations, we can rank these frameworks in terms of their overall performance: Android SQLite $\textgreater$ greenDAO $=$ DBFlow $\textgreater$ OrmLite $\textgreater$ Java Realm $\textgreater$ Paper $\textgreater$ Sugar ORM $\textgreater$ ActiveAndroid , where ``\textgreater" means ``having better performance than'', and ``='' means ``having similar performance with''.   
\item Considering the programming effort of implementing all database operations using different frameworks, DBFlow and Paper require less programming effort than the other frameworks. 
\item When considering the balance of programming effort and performance, DBFlow can be generally recommended for developing database-oriented mobile application with a standard database operation/schema complexity.  
\item Sugar ORM would not be an optimal choice when the dominating operations in a mobile app are select or delete, DBFlow would not be optimal when the dominating operation is insert, while Java Realm would not be optimal when the dominating operation is update. 
\end{enumerate}

\subsection{Experiments with the DaCapo benchmark}

\begin{figure*}[!ht]
\centering
\includegraphics[width=6in]{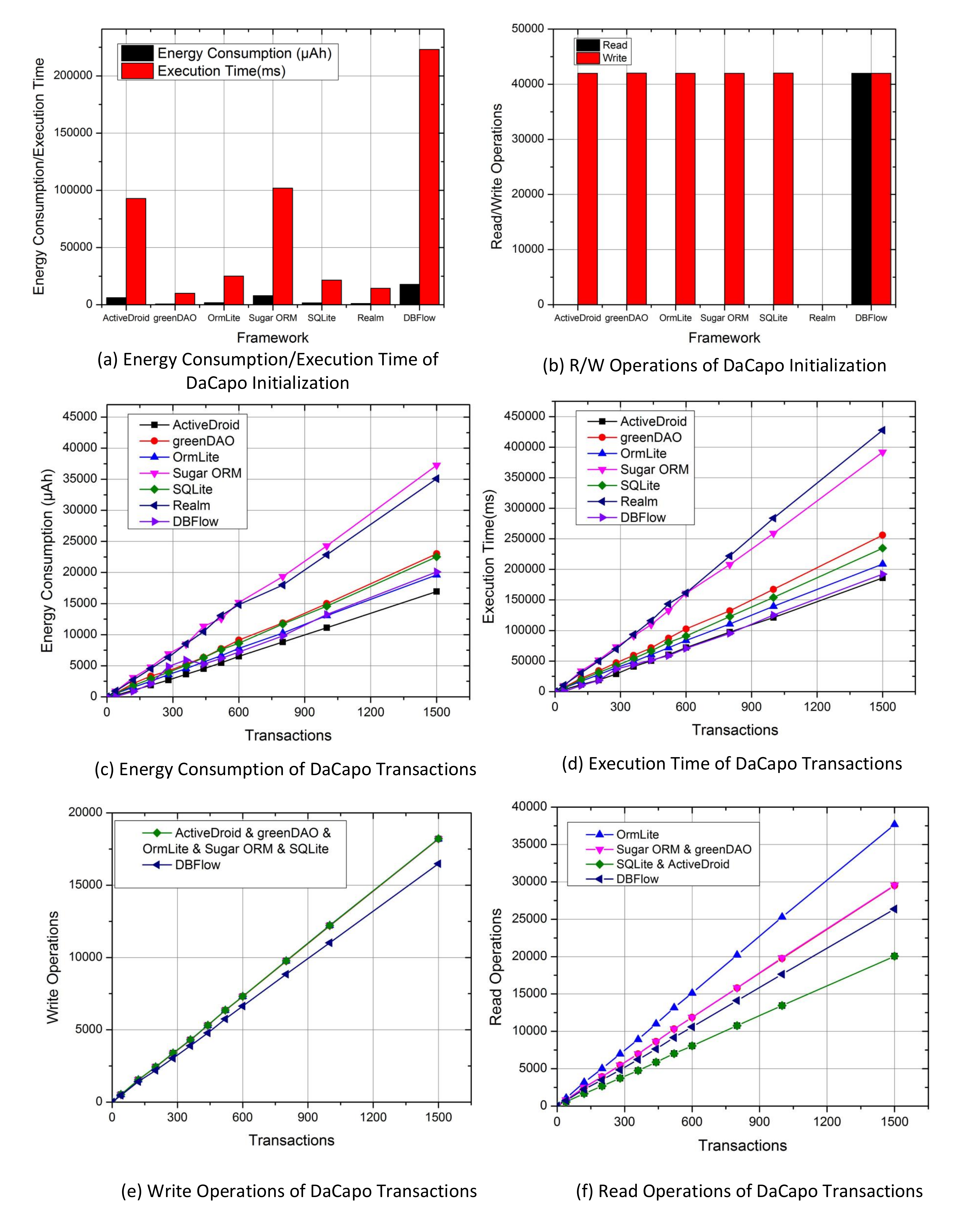}
\vspace{-3mm}
\caption{Energy/Execution time/Read and Write for DaCapo Benchmark with Alternative Persistence Frameworks}
\label{fig:dacapobenchmark}
\vspace{-3mm}
\end{figure*}

\begin{table*}[!ht]
\begin{center}
 \begin{tabular}{l|l}
Transaction Type & Operation Amount  \\ \hline
New order rollback & 8 \\ \hline
Order status by ID & 26 \\ \hline
Order status by name & 42   \\ \hline
Stock level & 59 \\ \hline
Delivery schedule & 62  \\ \hline
Payment by ID & 245  \\ \hline
Payment by name & 391  \\ \hline
New order & 667  \\ \hline

 \end{tabular}
\end{center}
\caption{Number of operations for each transaction type with 1500 overall transactions}
\label{table:dacapopa}
\vspace{-8mm}
\end{table*}

In this group of experiments, we use the DaCapo benchmark to study 
how the energy consumption and execution time of each framework changes in relation to the number of executed bank transactions.
The benchmark comes with a total of 41,971 records, so in our experiments we differ the number of bank transactions. In our measurements, we vary the number of bank transactions over the following values: 40, 120, 200, 280, 360, 440, 520, 600, 800, 1000, 1500. The total number of transactions is the sum of basic bank transactions, as listed in Table \ref{table:dacapopa}. Each transaction comprises a complex set of database operations. The key transactions in each run are ``New Order'', ``Payment by Name'', and ``Payment by ID'', which mainly execute the ``query" and ``update'' operations. In our experiments, ``New Order'' itself takes 42.5\% of the entire number of transactions. 

In \figurename \ref{fig:dacapobenchmark}, (a) and (b) show the energy/execution time and read/write operations of the DaCapo Initialization, respectively. \figurename \ref{fig:dacapobenchmark} (c) shows the energy consumption for each transaction number, and \figurename \ref{fig:dacapobenchmark} (d) shows the execution time for each transaction number. \figurename \ref{fig:dacapobenchmark} (e) and (f) show the read/write operation number, respectively. As the write operation number of ActiveDroid, greenDao, OrmLite, Sugar ORM and SQLite are very close (e.g, when the transaction amount is 1500, the write operation amount are 18209/18200/18205/18211/18212), we only present the average operation number of these five frameworks in \figurename \ref{fig:dacapobenchmark} (e). Similarly, we use the line in pink to present Sugar ORM/greenDao, and the line in green for SQLite/ActiveDroid in \figurename \ref{fig:dacapobenchmark} (f). Table \ref{table:dacapopa} shows the number of operations performed by each transaction.

The dominant database operation in the initialization phase is insert, and (a) shows the performance levels consistent with those seen in the Android ORM benchmark for the same operation: DBFlow, Sugar ORM and ActiveAndroid have the longest runtime. greenDAO performs better than Android SQLite, possibly due to greenDAO supporting batch insert (see Table \ref{table:featurecompare} for details). 

From \figurename \ref{fig:dacapobenchmark}, we observe that Java Realm and Sugar ORM have the longest execution time when executing the transactions whose major database operation is update (e.g., ``New order'', ``New order rollback'', ``Payment by name'', and ``Payment by ID''). This conclusion is consistent with that derived from the Android ORM update experiments shown in Section \ref{subsec:5a}. 
Android SQLite takes rather long to execute, as it involves database aggregation (e.g., $sum$, and the table queried had 30,060 records) and arithmetic operations (e.g. $field - 1$) in the select clause. Meanwhile, as ActiveAndroid only uses the raw SQL manipulation interface for complex update operations (Table \ref{table:featurecompare}), it exhibits the best performance, albeit at the cost of additional programming effort.

%Table \ref{table:dacapopa} also shows that greenDAO, ActiveAndroid, and Android SQLite incur higher execution costs for the ``Stock level'' transaction. One possible explanation is that this transaction contains a multiple entities conjunctive query action, and only these three frameworks provide the SQL ``JOIN" interface (Table. \ref{table:featurecompare}). Supporting this interface is known to be computationally expensive. However, the SQL ``JOIN'' interface can help save the programming effort. 

From \figurename \ref{fig:dacapobenchmark} (c-f) and Table \ref{table:dloc} we conclude that:
\begin{enumerate}[1]
\item ActiveAndroid offers the overall best performance for all DaCapo transactions. It shows the best performance for the most common transactions, at the cost of additional programming effort. Besides, its execution invokes the smallest number of database operations, due to its caching mechanism.
\item Sugar ORM and Java Realm have the longest execution time, in line with the Android ORM benchmark's results discussed in Section \ref{subsec:5a}. 
\item greenDAO's performance is in the middle, while requiring the lowest programming effort, taking 24.5\% fewer uncommented lines of code to implement than the other frameworks.
\item DBFlow takes more time and energy to execute that does ActiveAndroid; it also requires a higher programming effort than greenDAO does. Nevertheless, it strikes a good balance between the required programming effort and the resulting execution efficiency. 
\end{enumerate}

\subsection{Relationship of Energy Consumption, Execution Time and DB Operations}

\begin{figure*}[!ht]
\centering
\includegraphics[width=6.8in]{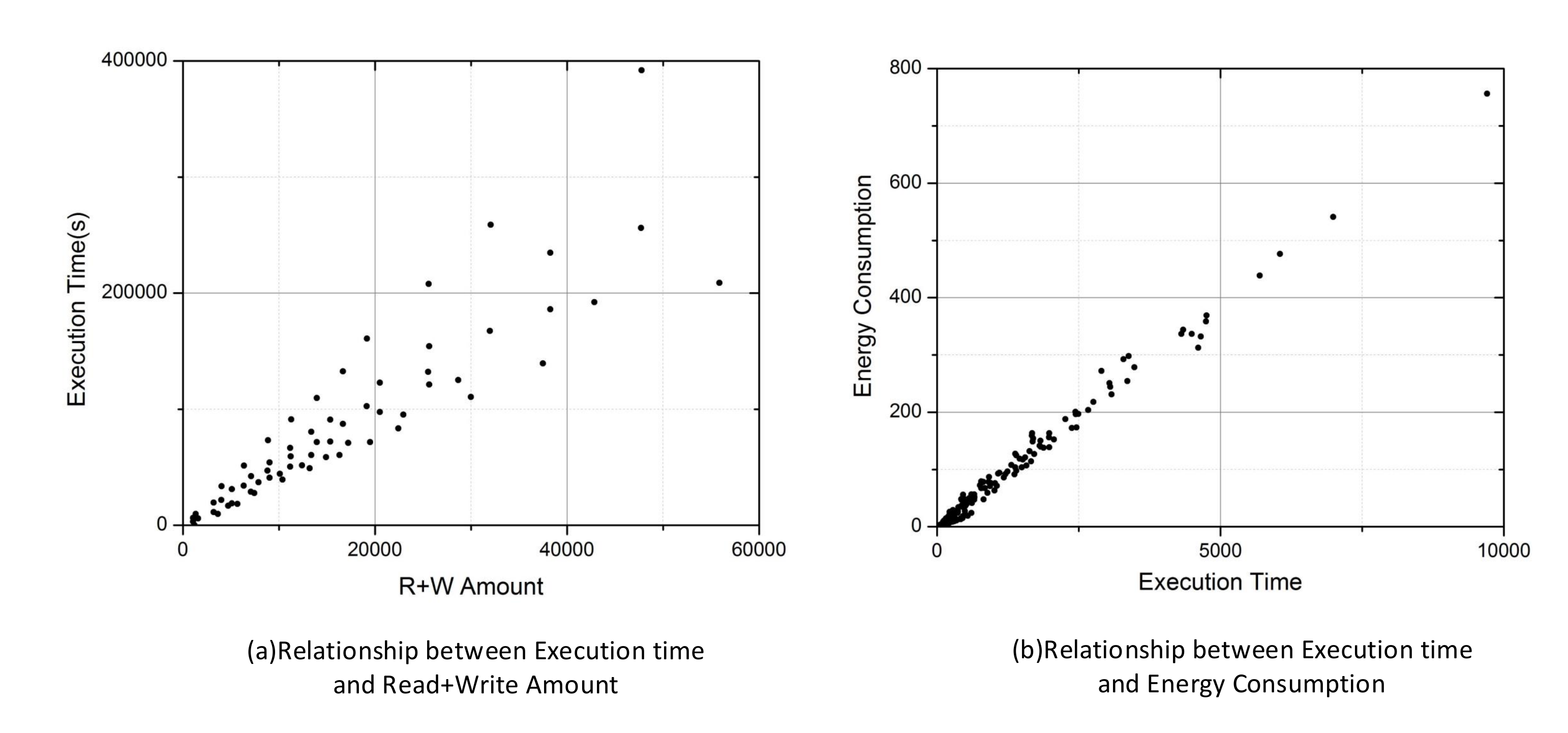}
\vspace{-3mm}
\caption{Energy/Execution Time/RW Amount Relationships}
\label{fig:relationships}
\vspace{-3mm}
\end{figure*}

We also discuss the relationship between energy consumption, execution time, and database operations.  We combine all the previous collected read/write, execution time and energy consumption data from all the benchmarks. As shown in \figurename \ref{fig:relationships} (a), the number of total database operations (Read and Write) has a dominating impact on both the execution time and energy consumption results: 1) as the number of data operations increases so usually do the execution time and energy consumption; 2) the execution time and energy consumption are also impacted by other factors (e.g., the complexity of data operations, the framework's implementation design choices, etc.). 

Meanwhile, as shown in Fig \ref{fig:relationships} (b),  there is a significant positive relationship between the time consumption and the energy consumption, with r(142) = 0.99, p$<$0.001. Hence, the longer a database task executes, the more energy it ends up consuming.

\begin{table}[!ht]
\begin{center}
 \begin{tabular}{lll}
 DaCapo & ULOC \\ \hline
 greenDAO & 2200 \\ \hline
 DBFlow & 2407  \\ \hline
 Sugar ORM & 2911 \\ \hline
 ActiveAndroid & 2923 \\ \hline
 Android SQLite & 3068 \\ \hline
 Java Realm & 3071  \\ \hline
 OrmLite & 3310 \\ \hline
 \end{tabular}
\end{center}
\caption{LOC for DaCapo Benchmark}
\label{table:dloc}
\vspace{-8mm}
\end{table}

\section{Numerical Model}
\label{sec:rec}
The experiments above show that for the same application scenario different frameworks exhibit different execution time, energy consumption, and programming effort. However, to derive practical benefits from these insights, mobile app developers need a way to quantify these trade-offs for any combination of an application scenario and a persistence framework. To that end, we propose a numerical model PEP (\textbf{P}erformance, \textbf{E}nergy consumption, and \textbf{P}rogramming \textbf{E}ffort) for mobile app developers to systematically evaluate the suitability of persistence frameworks. 
 
\subsection{Design of the PEP Model}
Our numerical model follows a commonly used technique for evaluating software products called \emph{a utility index}~\cite{christodoulou2009minimum}. Products with equivalent functionality possess multidimensional feature-sets, and a utility index is a score that quantifies the overall performance of each product, allowing the products to be compared with each other. 

As our experiments show, the utility of persistence frameworks is closely related to application features (e.g., data schema complexity, operations involved, data records manipulated, database operations executed). Therefore, it is only meaningful to compare the persistence frameworks within the context of a certain application scenario. Here, we use $p$ to denote an application with a set of features. 

%$\mathcal{O}=\{o=\textbf{ActiveAndroid},\textbf{greenDAO}, \textbf{OrmLite},\textbf{Sugar ORM},\textbf{Android SQLite},\textbf{Java Realm}, \textbf{DBFlow}, \textbf{Paper}\}$ 
Let $\mathcal{O}=\{o=1,2,3...\}$ be a set of frameworks. Let $E_o(p), \forall o \in \mathcal{O}$ denote the energy consumption of different implementations of $p$ using various frameworks $o$, while $T_o(p), \forall o \in \mathcal{O}$ denotes the execution time. As the energy consumption and the execution time of database operations correlate linearly in our experimental results, we use the Euclidean distance of a two dimensional vector to calculate the overall performance, which can be denoted as $P_o{p} = \sqrt{T_o(p)^2 + E_o(p)^2} , \forall o \in \mathcal{O}$. 

The programming effort is represented by the ULOC, and here we use $L_o(p), \forall o \in \mathcal{O}$ to denote the programming effort of different implementations of the project $p$ using different persistence frameworks. 

We consider both the framework's performance and programming effort to compute the utility index $I_o(p)$: 
\begin{equation}
I_o(p) = \frac{\min(P_o(p), \forall o \in \mathcal{O})}{P_o(p)} / \frac{L_o(p)}{\min(L_o(p), \forall o \in \mathcal{O})}, \forall o \in \mathcal{O}
\end{equation}

The equation's first part, $\frac{\min(P_o(p), \forall o \in \mathcal{O})}{P_o(p)}$, compares the performance of a mobile app implemented by means of the persistence framework $o$, and the implementation that has the best performance. The equation's second part, $\frac{L_o(p)}{\min(L_o(p), \forall o \in \mathcal{O})}$, compares the programming effort between an implementation $o(p)$ and the implementation that requires the minimal programming effort.  When the utility index of a framework $o$-based implementation is close to 1, the implementation is likely to offer acceptable performance, with low programming effort.

Consider the following example that demonstrates how to calculate the utility index. If for an app $p$, Android SQLite might provide the best performance, while the greenDAO-based implementation consumes twice the energy and takes twice the execution time. Therefore, the performance index of $P_{greenDAO}(p)$ is 0.5. On the other hand, the greenDAO-based implementation might require the lowest programming effort, as measured by the ULOC metric. Therefore, the implementation complexity index of $L_{greenDAO}{p}$ is 1. Thus, the overall utility index is 0.5/1 = 0.5. 

Application developers apply dissimilar standards to judge the trade-offs between performance and programming effort. Some developers focus solely on performance, while others may prefer the shortest time-to-market. We introduce $\tau$ to express these preferences. 

\begin{equation}
I_o(p) = \frac{\min(P_o(p), \forall o \in \mathcal{O})}{P_o(p)} / (\frac{L_o(p)}{\min(L_o(p), \forall o \in \mathcal{O})} )^\tau
\end{equation}

,where $ \tau > 0$. When $\tau >1$, the larger $\tau$ is, the more weight is assigned to the programming effort target. Otherwise, when $\tau<1$, the lower $\tau$ is, the more weight is assigned to the performance target. 

\subsection{Evaluating the Benchmarks}
To provide an insight into how the persistence frameworks evaluated in this article fit different  application scenarios, we apply the PEP model to the Android ORM and DaCapo benchmarks. We consider typical low and high transaction volumes, respectively, for each benchmark. Specifically, for the Android ORM benchmark, we evaluate two sets of input, 1,025 transactions and 20,025 transactions. For the DaCapo benchmark, we evaluate two sets of input, 40 transactions and 1,500 transactions. For each input set, we assign $\tau$ to 0.5, 1, and 1.5, in turn, to show whether the developers are willing to invest extra effort to improve performance. Specifically, when $\tau = 0.5$, the developer's main concern is performance; when $\tau = 1$, the balance of performance and programming effort is desired; when $\tau = 1.5$, the developer wishes to minimize programming effort. Table \ref{table:epperslt} shows the calculated index values of the persistence frameworks for all cases. 

From the results presented in Table \ref{table:epperslt}, we can draw several conclusions: 1) For the DaCapo benchmark or similar mobile apps with heavy data processing functionality and complicated data structures, DBFlow represents the best performance/programming effort trade-off. When the number of data operations is very high, ActiveAndroid should also be considered, as it provides the best execution performance, especially when the programmer is not as concerned about minimizing the programming effort; 2) For the Android ORM benchmark or similar mobile apps with less complicated data structures, when the number of data operations is small, the top choice is Paper, as this framework reduces the programming effort while providing high execution efficiency. However, when the number of data operations surpasses a certain threshold (over 10K simple data operations), the execution performance of Paper experiences a sudden drop due to scalability issues. In such cases, greenDAO and Android SQLite would be the recommended options. 

In the following discussion, we use hypothetical use cases to demonstrate how the generated guidelines can help mobile app developers pick the best framework for the application scenario in hand. Consider four cases: 1) a developer wants to persist the user's application-specific color scheme preferences; 2) a team of developers wants to develop a production-level contact book application; 3) an off-line map navigation application needs to store hundreds of MBs of data, comprising map fragments, points of interests, and navigation routines; 4) an MP3 player app needs to retrieve the artist's information based on some features of the MP3 being played. For use case 1, the main focus during the development procedure is to lower the programming effort, while the data structures and the number of data operations are simple and small. Therefore, for this use case, we would recommend using Paper. For use case 2, the main focus is to improve the responsiveness and efficiency, as the potential data volume can get quite large. Given that minimizing the programming effort is deprioritized, we would recommend using greenDAO or Android SQLite. For use case 3, complex data structures are required to be able to handle the potentially large data volumes, while maintaining quick responsiveness and high efficiency is expected of navigation apps. Therefore, we would recommend using ActiveAndroid. For use case 4, the main application's feature is playing MP3s, and the ability to retrieve the artist's data instantaneously is non-essential. To save the programming effort of this somewhat auxiliary feature, we would recommend using DBFlow. 

\begin{table*}[!ht]
\scriptsize
\begin{center}
\begin{tabular}{l|l|l|l|l|l|l|l|l|l}
\toprule
\textbf{Benchmark} & \textbf{Parameters \& Index} & \textbf{ActiveAndroid} & \textbf{greenDAO} & \textbf{OrmLite} & \textbf{Sugar} & \textbf{SQLite} & \textbf{Realm} & \textbf{DBFlow} & \textbf{Paper} \\ \hline
DaCapo & trans=40,$\tau$=0.5 & 0.328 & 0.184 & 0.135 & 0.166 & 0.109 & 0.098 &  \textbf{0.956} & N.A. \\ \hline
DaCapo & trans=40,$\tau$=1 & 0.284 & 0.184 & 0.135 & 0.094 & 0.140 & 0.083 & \textbf{0.914} & N.A.\\ \hline
DaCapo & trans=40,$\tau$=1.5 & 0.247 & 0.184 & 0110 & 0.082 & 0.118 & 0.070 & \textbf{0.878} & N.A. \\ \hline
DaCapo & trans=1500,$\tau$=0.5 & \textbf{0.867} & 0.726 & 0.726 & 0.412 & 0.671 & 0.368 & \textbf{0.924} & N.A.  \\ \hline
DaCapo & trans=1500,$\tau$=1 &  0.752 & 0.726 & 0.592 & 0.358 & 0.568 & 0.312 & \textbf{0.884} & N.A. \\ \hline
DaCapo & trans=1500,$\tau$=1.5 &  0.652 & 0.726 & 0.483 & 0.133 & 0.481 & 0.264 & \textbf{0.845} & N.A \\ \hline\hline
Android ORM & trans=1025,$\tau$=0.5 & 0.050 & 0.166 & 0.124 & 0.044 & 0.180 & 0.055 & 0.255 & \textbf{0.976} \\ \hline
Android ORM & trans=1025,$\tau$=1 & 0.051 & 0.144 & 0.092 & 0.039 & 0.139 & 0.042 & 0.255 & \textbf{0.952} \\ \hline
Android ORM & trans=1025,$\tau$=1.5 & 0.043 & 0.125 & 0.069 & 0.035 & 0.107 & 0.031 & 0.255 & \textbf{0.929} \\ \hline
Android ORM & trans=20025,$\tau$=0.5 & 0.159 & \textbf{0.740} & 0.633 & 0.289 & \textbf{0.769} & 0.007 & 0.591 & 0.635  \\ \hline
Android ORM & trans=20025,$\tau$=1 	& 0.165 & \textbf{0.641} & 0.472 & 0.080 & 0.591 & 0.005 & 0.591 & 0.621 \\ \hline
Android ORM & trans=20025,$\tau$=1.5 &  0.114 & 0.555 & 0.352 & 0.071 & 0.454 & 0.004 & 0.591 & \textbf{0.606}	\\ \hline
\end{tabular}
\end{center}
\caption{Applying Numerical Model on DaCapo and Android ORM Benchmark}
\label{table:epperslt}
\vspace{-8mm}
\end{table*}

\section{Threats to Validity}
\label{sec:threats}
Next, we discuss the threats to the validity of our experimental results. Although in designing our experimental evaluation, we tried to perform as an objective assessment as possible, our design choices could have certainly affected the validity and applicability of our conclusions.

The key external threat to validity is our choice of the hardware devices, Android version, and profiling equipment. Specifically,
we conduct our experiments with an LG mobile phone, with 1.2GHz quad-core Qualcomm Snapdragon 400 processor, running Android 4.4.2 KitKat, profiled with the Monsoon Power Monitor. Even though these experimental parameters are representative of the Android computing ecosystem, changing any of these parameters could have affected some outcomes of our experiments. 

The key internal threat to validity are our design choices for structuring the database and the persistence application functionality. Specifically, while our Android ORM benchmark set explores the object features of Android persistence frameworks, the original DaCapo \cite{blackburn2006dacapo} H2 benchmark manipulates relational database structures directly, without stress-testing the object-oriented persistence frameworks around it. To retarget DaCapo to focus on persistence frameworks rather than the JVM alone, we adapted the benchmark to make use of transparent persistence as a means of accessing its database-related functionality. Nevertheless, the relatively large scale of data volume, with the select and update operations bank transactions dominating the execution, this benchmark is representative of a large class of database application systems, but not all of them. Besides, we have not tested our PEP model on real-world applications. Hence, it is not confirmed yet how accurate the model would be for such applications. 

\section{Conclusions}
\label{sec:conclusion}
In this paper, we present a systematic study of popular Android ORM/OO persistence frameworks. We first compare and contrast the frameworks to present an overview of their features and capabilities. Then we present our experimental design of two sets of benchmarks, used to explore the execution time, energy consumption, and programming effort of these frameworks in different application scenarios. We analyze our experimental results in the context of the analyzed frameworks' features and capabilities. Finally, we propose a numerical model to help guide mobile developers in their decision making process when choosing a persistence framework for a given application scenario. To the best of our knowledge, this research is the first step to better understand the trade-offs between the execution time, energy efficiency, and programming effort of Android persistence frameworks. As a future work direction, we plan to apply the PEP model presented above to real-world applications, in order to assess its accuracy and applicability.


\section*{Acknowledgment}
This research is supported by the National Science Foundation
through the Grants CCF-1649583 and CCF-1717065.



\balance
\bibliographystyle{IEEEtran}
\bibliography{persistencebib.bib}

\begin{thebibliography}{10}
\providecommand{\url}[1]{#1}
\csname url@samestyle\endcsname
\providecommand{\newblock}{\relax}
\providecommand{\bibinfo}[2]{#2}
\providecommand{\BIBentrySTDinterwordspacing}{\spaceskip=0pt\relax}
\providecommand{\BIBentryALTinterwordstretchfactor}{4}
\providecommand{\BIBentryALTinterwordspacing}{\spaceskip=\fontdimen2\font plus
\BIBentryALTinterwordstretchfactor\fontdimen3\font minus
  \fontdimen4\font\relax}
\providecommand{\BIBforeignlanguage}[2]{{%
\expandafter\ifx\csname l@#1\endcsname\relax
\typeout{** WARNING: IEEEtran.bst: No hyphenation pattern has been}%
\typeout{** loaded for the language `#1'. Using the pattern for}%
\typeout{** the default language instead.}%
\else
\language=\csname l@#1\endcsname
\fi
#2}}
\providecommand{\BIBdecl}{\relax}
\BIBdecl

\bibitem{Androidshare2015Q2}
\BIBentryALTinterwordspacing
``Smartphone {OS} market share,'' 2015, Last accessed data: 19-June-2018.
  [Online]. Available:
  \url{http://www.idc.com/prodserv/smartphone-os-market-share.jsp}
\BIBentrySTDinterwordspacing

\bibitem{Androidapps2015July}
\BIBentryALTinterwordspacing
``Statista: Mobile apps available in leading stores,'' 2015, Last accessed
  data: 19-June-2018. [Online]. Available:
  \url{http://www.statista.com/statistics/276623/number-of-apps-available-in-leading-app-stores/}
\BIBentrySTDinterwordspacing

\bibitem{jha2011poorly}
S.~Jha, ``Poorly written apps can sap 30 to 40\% of a phone’s juice,'' in
  \emph{CEO, Motorola Mobility, Bank of America Merrill Lynch 2011 Technology
  Conference}, June 2011.

\bibitem{kwon2013impact}
Y.-W. Kwon and E.~Tilevich, ``The impact of distributed programming
  abstractions on application energy consumption,'' \emph{Information and
  Software Technology}, vol.~55, no.~9, pp. 1602--1613, 2013.

\bibitem{li2014empirical}
D.~Li, S.~Hao, J.~Gui, and W.~G. Halfond, ``An empirical study of the energy
  consumption of {A}ndroid applications,'' in \emph{Software Maintenance and
  Evolution (ICSME), 2014 IEEE International Conference on}.\hskip 1em plus
  0.5em minus 0.4em\relax IEEE, 2014, pp. 121--130.

\bibitem{hao2013estimating}
S.~Hao, D.~Li, W.~G. Halfond, and R.~Govindan, ``Estimating mobile application
  energy consumption using program analysis,'' in \emph{Software Engineering
  (ICSE), 2013 35th International Conference on}.\hskip 1em plus 0.5em minus
  0.4em\relax IEEE, 2013, pp. 92--101.

\bibitem{tiwari1996instruction}
V.~Tiwari, S.~Malik, A.~Wolfe, and M.~T.-C. Lee, ``Instruction level power
  analysis and optimization of software,'' in \emph{Technologies for wireless
  computing}.\hskip 1em plus 0.5em minus 0.4em\relax Springer, 1996, pp.
  139--154.

\bibitem{dong2010sesame}
M.~Dong and L.~Zhong, ``Sesame: Self-constructive system energy modeling for
  battery-powered mobile systems,'' \emph{arXiv preprint arXiv:1012.2831},
  2010.

\bibitem{cohen2012energy}
M.~Cohen, H.~S. Zhu, E.~E. Senem, and Y.~D. Liu, ``Energy types,'' in \emph{ACM
  SIGPLAN Notices}, vol.~47, no.~10.\hskip 1em plus 0.5em minus 0.4em\relax
  ACM, 2012, pp. 831--850.

\bibitem{sahin2012initial}
C.~Sahin, F.~Cayci, I.~L.~M. Guti{\'e}rrez, J.~Clause, F.~Kiamilev, L.~Pollock,
  and K.~Winbladh, ``Initial explorations on design pattern energy usage,'' in
  \emph{Green and Sustainable Software (GREENS), 2012 First International
  Workshop on}.\hskip 1em plus 0.5em minus 0.4em\relax IEEE, 2012, pp. 55--61.

\bibitem{hindle2012green}
A.~Hindle, ``Green mining: A methodology of relating software change to power
  consumption,'' in \emph{Proceedings of the 9th IEEE Working Conference on
  Mining Software Repositories}.\hskip 1em plus 0.5em minus 0.4em\relax IEEE
  Press, 2012, pp. 78--87.

\bibitem{kwon2013reducing}
Y.-W. Kwon and E.~Tilevich, ``Reducing the energy consumption of mobile
  applications behind the scenes,'' in \emph{2013 IEEE International Conference
  on Software Maintenance}.\hskip 1em plus 0.5em minus 0.4em\relax IEEE, 2013,
  pp. 170--179.

\bibitem{pinto2014mining}
G.~Pinto, F.~Castor, and Y.~D. Liu, ``Mining questions about software energy
  consumption,'' in \emph{Proceedings of the 11th Working Conference on Mining
  Software Repositories}.\hskip 1em plus 0.5em minus 0.4em\relax ACM, 2014, pp.
  22--31.

\bibitem{mukherjee2014future}
S.~Mukherjee and I.~Mondal, ``Future practicability of {A}ndroid application
  development with new {A}ndroid libraries and frameworks,''
  \emph{International Journal of Computer Science and Information
  Technologies}, vol.~5, no.~4, pp. 5575--5579, 2014.

\bibitem{activeandroid}
\BIBentryALTinterwordspacing
``Activeandroid by pardom,'' Last accessed data: 19-June-2018. [Online].
  Available: \url{http://www.activeandroid.com/}
\BIBentrySTDinterwordspacing

\bibitem{greendao}
\BIBentryALTinterwordspacing
``{greenDAO}: the superfast android orm for sqlite,'' Last accessed data:
  19-June-2018. [Online]. Available: \url{http://greenrobot.org/greendao/}
\BIBentrySTDinterwordspacing

\bibitem{ormlite}
\BIBentryALTinterwordspacing
``{OrmLite} -- lightweight object relational mapping ({ORM}) {J}ava package,''
  Last accessed data: 19-June-2018. [Online]. Available:
  \url{http://greenrobot.org/greendao/}
\BIBentrySTDinterwordspacing

\bibitem{sugarorm}
\BIBentryALTinterwordspacing
``{Sugar ORM} -- insanely easy way to work with {A}ndroid databases,'' Last
  accessed data: 19-June-2018. [Online]. Available:
  \url{http://satyan.github.io/sugar/}
\BIBentrySTDinterwordspacing

\bibitem{sqlite}
\BIBentryALTinterwordspacing
``Sqlite database engine,'' Last accessed data: 19-June-2018. [Online].
  Available: \url{https://www.sqlite.org/}
\BIBentrySTDinterwordspacing

\bibitem{DBFlow}
\BIBentryALTinterwordspacing
``{DBFlow}: A blazing fast, powerful, and very simple {ORM} {A}ndroid database
  library that writes database code for you.'' Last accessed data:
  19-June-2018. [Online]. Available: \url{https://github.com/Raizlabs/DBFlow}
\BIBentrySTDinterwordspacing

\bibitem{realm}
\BIBentryALTinterwordspacing
``Realm database engine,'' Last accessed data: 19-June-2018. [Online].
  Available: \url{https://realm.io/}
\BIBentrySTDinterwordspacing

\bibitem{Paper}
\BIBentryALTinterwordspacing
``Paper: Fast and simple data storage library for {A}ndroid.'' Last accessed
  data: 19-June-2018. [Online]. Available: \url{https://github.com/pilgr/Paper}
\BIBentrySTDinterwordspacing

\bibitem{blackburn2006dacapo}
S.~M. Blackburn, R.~Garner, C.~Hoffmann, A.~M. Khang, K.~S. McKinley,
  R.~Bentzur, A.~Diwan, D.~Feinberg, D.~Frampton, S.~Z. Guyer \emph{et~al.},
  ``The {DaCapo} benchmarks: {J}ava benchmarking development and analysis,'' in
  \emph{ACM Sigplan Notices}, vol.~41, no.~10.\hskip 1em plus 0.5em minus
  0.4em\relax ACM, 2006, pp. 169--190.

\bibitem{jing16}
P.~Jing, S.~Zheng, and T.~Eli, ``Understanding the energy, performance, and
  programming effort trade-offs of {A}ndroid persistence frameworks,'' in
  \emph{MASCOTS'16}.\hskip 1em plus 0.5em minus 0.4em\relax IEEE, 2016.

\bibitem{banerjee2014detecting}
A.~Banerjee, L.~K. Chong, S.~Chattopadhyay, and A.~Roychoudhury, ``Detecting
  energy bugs and hotspots in mobile apps,'' in \emph{Proceedings of the 22nd
  ACM SIGSOFT International Symposium on Foundations of Software
  Engineering}.\hskip 1em plus 0.5em minus 0.4em\relax ACM, 2014, pp. 588--598.

\bibitem{chowdhury2017exploratory}
S.~Chowdhury, S.~Di~Nardo, A.~Hindle, and Z.~M.~J. Jiang, ``An exploratory
  study on assessing the energy impact of logging on {A}ndroid applications,''
  \emph{Empirical Software Engineering}, pp. 1--35, 2017.

\bibitem{chowdhury2016client}
S.~A. Chowdhury, V.~Sapra, and A.~Hindle, ``Client-side energy efficiency of
  {HTTP}/2 for web and mobile app developers,'' in \emph{Software Analysis,
  Evolution, and Reengineering (SANER), 2016 IEEE 23rd International Conference
  on}, vol.~1.\hskip 1em plus 0.5em minus 0.4em\relax IEEE, 2016, pp. 529--540.

\bibitem{liu2016understanding}
Y.~Liu, C.~Xu, S.-C. Cheung, and V.~Terragni, ``Understanding and detecting
  wake lock misuses for {A}ndroid applications,'' in \emph{Proceedings of the
  2016 24th ACM SIGSOFT International Symposium on Foundations of Software
  Engineering}.\hskip 1em plus 0.5em minus 0.4em\relax ACM, 2016, pp. 396--409.

\bibitem{hindle2014greenminer}
A.~Hindle, A.~Wilson, K.~Rasmussen, E.~J. Barlow, J.~C. Campbell, and
  S.~Romansky, ``Greenminer: A hardware based mining software repositories
  software energy consumption framework,'' in \emph{Proceedings of the 11th
  Working Conference on Mining Software Repositories}.\hskip 1em plus 0.5em
  minus 0.4em\relax ACM, 2014, pp. 12--21.

\bibitem{monsoon}
\BIBentryALTinterwordspacing
``Monsoon power monitor,'' Last accessed data: 19-June-2018. [Online].
  Available: \url{https://www.msoon.com/LabEquipment/PowerMonitor/}
\BIBentrySTDinterwordspacing

\bibitem{linares2014mining}
M.~Linares-V{\'a}squez, G.~Bavota, C.~Bernal-C{\'a}rdenas, R.~Oliveto,
  M.~Di~Penta, and D.~Poshyvanyk, ``Mining energy-greedy {API} usage patterns
  in {A}ndroid apps: an empirical study,'' in \emph{Proceedings of the 11th
  Working Conference on Mining Software Repositories}.\hskip 1em plus 0.5em
  minus 0.4em\relax ACM, 2014, pp. 2--11.

\bibitem{hasan2016energy}
S.~Hasan, Z.~King, M.~Hafiz, M.~Sayagh, B.~Adams, and A.~Hindle, ``Energy
  profiles of {J}ava collections classes,'' in \emph{Software Engineering
  (ICSE), 2016 IEEE/ACM 38th International Conference on}.\hskip 1em plus 0.5em
  minus 0.4em\relax IEEE, 2016, pp. 225--236.

\bibitem{pathak2012energy}
A.~Pathak, Y.~C. Hu, and M.~Zhang, ``Where is the energy spent inside my app?:
  fine grained energy accounting on smartphones with eprof,'' in
  \emph{Proceedings of the 7th ACM european conference on Computer
  Systems}.\hskip 1em plus 0.5em minus 0.4em\relax ACM, 2012, pp. 29--42.

\bibitem{li2016automated}
D.~Li, Y.~Lyu, J.~Gui, and W.~G. Halfond, ``Automated energy optimization of
  {HTTP} requests for mobile applications,'' in \emph{Software Engineering
  (ICSE), 2016 IEEE/ACM 38th International Conference on}.\hskip 1em plus 0.5em
  minus 0.4em\relax IEEE, 2016, pp. 249--260.

\bibitem{manotas2016empirical}
I.~Manotas, C.~Bird, R.~Zhang, D.~Shepherd, C.~Jaspan, C.~Sadowski, L.~Pollock,
  and J.~Clause, ``An empirical study of practitioners' perspectives on green
  software engineering,'' in \emph{Software Engineering (ICSE), 2016 IEEE/ACM
  38th International Conference on}.\hskip 1em plus 0.5em minus 0.4em\relax
  IEEE, 2016, pp. 237--248.

\bibitem{lyu2017empirical}
Y.~Lyu, J.~Gui, M.~Wan, and W.~G. Halfond, ``An empirical study of local
  database usage in {A}ndroid applications,'' in \emph{Software Maintenance and
  Evolution (ICSME), 2017 IEEE International Conference on}.\hskip 1em plus
  0.5em minus 0.4em\relax IEEE, 2017, pp. 444--455.

\bibitem{xia2009efficient}
C.~Xia, G.~Yu, and M.~Tang, ``Efficient implement of {ORM} (object/relational
  mapping) use in {J2EE} framework: {H}ibernate,'' in \emph{Computational
  Intelligence and Software Engineering, 2009. CiSE 2009. International
  Conference on}.\hskip 1em plus 0.5em minus 0.4em\relax IEEE, 2009, pp. 1--3.

\bibitem{cvetkovic2010comparative}
S.~Cvetkovi{\'c} and D.~Jankovi{\'c}, ``A comparative study of the features and
  performance of {ORM} tools in a .{NET} environment,'' in \emph{Objects and
  Databases}.\hskip 1em plus 0.5em minus 0.4em\relax Springer, 2010, pp.
  147--158.

\bibitem{vetro2013definition}
A.~Vetro, L.~Ardito, G.~Procaccianti, and M.~Morisio, ``Definition,
  implementation and validation of energy code smells: an exploratory study on
  an embedded system,'' 2013.

\bibitem{subramanian1999performance}
M.~Subramanian, V.~Krishnamurthy, and R.~Shores, ``Performance challenges in
  object-relational {DBMS}s,'' \emph{IEEE Data Eng. Bull.}, vol.~22, no.~2, pp.
  27--31, 1999.

\bibitem{newman2004sqlite}
C.~Newman, \emph{{SQLite} (Developer's Library)}.\hskip 1em plus 0.5em minus
  0.4em\relax Sams, 2004.

\bibitem{NoSQL}
\BIBentryALTinterwordspacing
``{NoSQL} database engine,'' Last accessed data: 19-June-2018. [Online].
  Available: \url{http://nosql-database.org/}
\BIBentrySTDinterwordspacing

\bibitem{lavazza2016empirical}
L.~Lavazza, S.~Morasca, and D.~Tosi, ``An empirical study on the effect of
  programming languages on productivity,'' in \emph{Proceedings of the 31st
  Annual ACM Symposium on Applied Computing}.\hskip 1em plus 0.5em minus
  0.4em\relax ACM, 2016, pp. 1434--1439.

\bibitem{singh2014evolution}
M.~P. Singh and M.~K. Jain, ``Evolution of processor architecture in mobile
  phones,'' \emph{International Journal of Computer Applications}, vol.~90,
  no.~4, 2014.

\bibitem{christodoulou2009minimum}
S.~E. Christodoulou, G.~Ellinas, and A.~Michaelidou-Kamenou, ``Minimum moment
  method for resource leveling using entropy maximization,'' \emph{Journal of
  construction engineering and management}, vol. 136, no.~5, pp. 518--527,
  2009.

\end{thebibliography}
%\begin{thebibliography}{1}

%H.~Kopka and P.~W. Daly, \emph{A Guide to \LaTeX}, %3rd~ed.\hskip 1em plus
%  0.5em minus 0.4em\relax Harlow, England: Addison-Wesley, %1999.

%\end{thebibliography}

\end{document}